# Transient Stability Analysis for Grid Following Converters in Low-Inertia Power Systems by Direct Method

Fangyuan Sun[1], Ruisheng Diao[1]*, Ruiyuan Zeng[1], Zhanning Liu[1], Baorong Zhou[2], Junjie Li[2], Wangqianyun Tang[2]

(1. ZJU-UIUC Institute, Zhejiang University, Haining, 314400 China
2. China Southern Grid Electric Power Research Institute Co., Ltd, Guangzhou, 510623 China)

*Abstract*—**With the increased penetration of renewable energy and reduced proportion of synchronous generators, the low-inertia characteristics of today's power system become prominent, and the transient stability issue of grid following converter (GFLC) under low inertia system (LIS) condition becomes critical. There are two prominent problems in the transient stability analysis of GFLC-LIS: The angular dynamic of LIS increases the complexity of transient stability analysis, and the nonlinear, possibly negative damping of GFLC makes it difficult to guarantee the conservative of the traditional methods. These problems make the traditional methods inapplicable. In this paper, the transient stability analysis of GFLC-LIS is investigated to provide an accurate estimation of the attraction boundary and critical clearance time (CCT). Firstly, a dynamic model of GFLC-LIS is constructed, considering the phase-locked loop (PLL)-based GFLC dynamics and swing equation-based LIS dynamics. The frequency mutation of PLL at fault occurrence and clearing time is also considered. Secondly, a Zubov-based transient stability analysis method is proposed, which can construct the energy function in a way that is different from the traditional conservation of energy perspective and can address the negative damping issue. Moreover, the accuracy of the CCT estimation is analyzed, and the influences of LIS parameters on transient stability are illustrated. Finally, simulation experiments are carried out to verify the effectiveness of the proposed method.**

## I. INTRODUCTION

WITH the ever-growing penetration of renewable energy into modern bulk power systems, the proportion of conventional synchronous generators (SGs) is declining, with inverter-based generation (IBG) gradually becoming the dominating power supply [1]. Grid following converters (GFLCs) are the main interface for renewable energy integration, in which the phase-locked loop (PLL)-based mechanism for synchronization is significantly different from that of SGs, causing new transient behavior under major disturbances. Thus, transient stability analysis of GFLC has become an important research topic. With a reduced proportion of SGs that traditionally provide both frequency and voltage support, the power system has been gradually weakened, which significantly influences the transient stability of the system with GFLCs [2] [3].

The traditional definition of weak power systems generally refers to the weak voltage support due to long-distance transmission lines or heavy load conditions. However, with the increased penetration of inverter-based resources, the inertia problem also becomes prominent, which makes the system-side power angle dynamics more sensitive under large disturbances, thereby affecting transient synchronization stability [4]. Nowadays, the weakness of a power system can be reflected in two aspects, including a lack of voltage support and a lack of inertia support [5]. The former indicates that the system has a small voltage influence on the point of common coupling (PCC), which is manifested as a large short circuit ratio (SCR) [5]. The stability of a high-IBG-penetration system with weak voltage support has been widely studied [6][7]. The current research on the transient stability of GFLC mainly focuses on the influence of SCR and fault ride-through enhancement methods [8][9][10], but the definition of the weak system is generally limited to weak voltage support. Lack of inertia is another problem caused by decreased SG proportion [12]. Larger frequency fluctuation in power systems significantly influences the interaction with IBRs, but most of the relevant researches focus on small-signal stability only. In [13] and [14], the small-signal stability of GFLC connecting to a low-inertia system (LIS) is studied, with key control loops analyzed. The influence of LIS on the transient stability of GFLC is studied in [4], but only the sustained fault is considered in the stability analysis, and fault clearing is not considered.

Traditional direct methods for transient analysis of power systems are highly related to the transient behavior of SGs and other equipment. The Equal Area Criterion (EAC) and Lyapunov method are the most widely used methods. Current research efforts mainly focus on the application of traditional methods on IBR-based bulk power systems [15][16]. It is reported that the dynamic equation of GFLC is similar in form to the swing equation of SG [17][18], but the negative damping issue under certain phase angle ranges brings difficulties to the application of traditional methods in assessing transient stability. The traditional transient analysis methods are mainly based on the conservation of energy perspective, and the positive damping characteristics of SGs make the damping term negligible [19]. However, the potentially negative damping of GFLC is non-ignorable. Calculating energy consumption/accumulation of damping requires a post-fault trajectory (i.e., path-dependent), which means post-fault simulation is required to get the value of variables at each time step, thus losing the time-saving advantage compared to the numerical simulation-based methods. In [20], the EAC method is used for the transient stability assessment of GFLC, but the damping term is neglected. The stability analysis is implemented under various contingencies, and results show that in some cases, the negative damping can lead to radical estimation of critical clearance time (CCT). In [21], the energy dissipation/accumulation of damping is considered in the proposed iterative EAC method. The calculation of damping energy is based on the iterative approximation of the fault trajectory. However, only the first swing of the fault trajectory is considered in this approximation, but the negative damping may lead to desynchronization during the second swing, which is not considered. In [22], the Lyapunov method is used to estimate the stability boundary of the GFLC. However, to satisfy the requirements of the energy function, the positive damping range is regarded as the attraction domain, which leads to a large conservative error. In [23][24], port-Hamilton theory is used for transient stability, but positive damping is also a requirement of the estimated attraction domain, and a large conservative error is inevitable. In [4], the Lyapunov method is used for transient stability analysis of a GFLC-LIS. To neglect the positive damping requirement, the energy dissipation in the system under sustained fault is proved, but fault clearing is not considered. Approximation methods like ray approximation [25] or trapezoid approximation [26] can estimate the energy dissipation/accumulation of damping, but radical error can be introduced. To the authors' best knowledge, these approximation methods have not been well applied to the study of GFLC transient stability. In summary, two ways are used to solve the abovementioned negative damping problem, including 1) Limiting the attraction domain in the positive damping region, which leads to a large conservative error, and 2) Calculating or analyzing the path-dependent energy dissipation/accumulation of damping. This is feasible under sustained fault, but the fault trajectory is significantly more complex when fault clearing is considered.

The widely used GFLC connected to an ideal voltage source model is considered a 2-dimensional model with two variables and two differential equations. When the angular dynamics of LIS are considered, new variables and differential equations are introduced to GFLC-LIS model, i.e., the dimension of the model is increased. The traditional EAC method is generally used for a 2-dimensional model, which makes it inapplicable for the GFLC-LIS model. Reducing the GFLC-LIS model dimension to two will reduce the accuracy of the model. The Lyapunov method is applicable, but the traditional energy function composed of kinetic energy, potential energy, and damping energy dissipation does not meet the Lyapunov requirements because of the negative damping. The inability of traditional transient analysis methods is further illustrated in Section III, according to the dynamic equation of the GFLC-LIS.

In summary, 1) the negative damping problem results in the invalidity of the conventional energy function. 2) The newly introduced angular dynamic of ILS increases dimensionality, thereby rendering the EAC method inapplicable, and the radical error introduced by negative damping issues also persists.

To address the problem mentioned above, it is worth exploring the application of stability region evaluation methods other than traditional methods. The Zubov method provides an effective means for constructing the Lyapunov function. Different from the perspective of conservation of energy, Zubov can determine the energy function of a dynamic system with polynomial form by recursion[27]. The construction of the energy function is more time-consuming compared to the traditional method, but the energy function and attraction domain are valid for any faults, and the post-fault trajectory analysis is not required. In [28], the Zubov method is used for transient analysis of salient pole synchronous generator, but it has not been used for IBR stability analysis to the authors' best knowledge.

In this paper, a Zubov-based transient analysis method for the GFLC-LIS is proposed in this paper. The angular dynamics of GFLC are considered in the model, and the energy function considering potentially negative damping is constructed. The main contributions of this paper are summarized as follows:

1) A dynamic model for GFLC-LIS is established. Both angular dynamics of LIS and frequency mutation of GFLCs are considered.

2) The energy function for the proposed model is constructed using the Zubov method. Different from the traditional energy function from the energy conservation perspective, the new energy function can guarantee the conservative of the estimation in negatively damped regions, without requiring post-fault simulation.

3) The attraction domain of GFLC-LIS is estimated. The correctness and accuracy of the proposed method are verified by simulation results, and the influences of Zubov method parameters and LIS parameters on transient stability are analyzed.

The remainder of the paper is organized as follows. Section II introduces the dynamic model of a GFLC-LIS. In Section III, a detailed transient stability analysis approach based on the Zubov method is given, and the impacts of parameters are analyzed. Simulation and test results are given in Section IV. Eventually, Section V draws the conclusion.

## II. SYSTEM MODELING

The traditional equivalence of the GFLC-connected system is based on the single-machine-infinite-system model, which uses the impedance between the GFLC and the voltage source to reflect the voltage support of the system. However, the angular dynamics of the low-inertia system cannot be reflected by the voltage source. In this section, a two-machine system is used to simulate the GFLC connected to a low-inertia system scenario. The system diagram is shown in Fig.1, which consists of one converter, one SG, and four impedances. The following assumptions are made throughout this paper:

1) The low-inertia system dynamics are expressed by the swing equation of the SG [4],[29]. The phase angle is represented by $\delta_g$. Under a large disturbance, the angular dynamics of the LIS can be represented by the SG dynamics.

2) The voltage support to the GFLC is reflected by the impedance $Z_c$ and $Z_g$ between the SG and GFLC. Low voltage support corresponds to large connection impedance. So, SG only needs to simulate the angular dynamics of the system, and the terminal voltage $U_g$ is considered constant.

3) The current control loop of GFLC is neglected due to its short timescale. The GFLC can be regarded as a controlled current source with fixed current amplitude, $I_c$, and the phase angle $\delta_c$ is determined by the phase-locked loop (PLL).

4) The load is modeled as a constant impedance $Z_l$. When a three-fault short-circuit fault occurs, the fault resistance $R_f$ is connected in parallel with $Z_l$.

5) The frequency fluctuation is generally small in the bulk power system transient processes; its influence on the reactance is ignored, and the reactance in the system is regarded as constant.

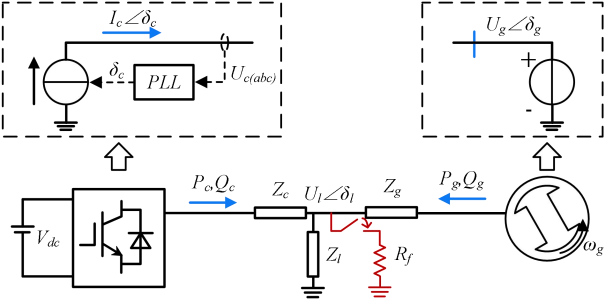

**Fig.1.** Diagram of GFLC-LIS

All the dynamics are modeled under DQ frame. Three DQ frames, $d_0$-$q_0$, $d_g$-$q_g$, and $d_c$-$q_c$ are used to describe the variables of GFLC-LIS which are shown in Fig.2. $d_0$-$q_0$ frame is the reference frame with a constant angular velocity $\omega_0$ ($120\pi$ in this paper). $d_g$-$q_g$ and $d_c$-$q_c$ frame are frames of LIS and GFLC with variable angular velocity $\omega_g^t$, and $\omega_c^t$. Their phase differences from $d_0$-$q_0$ frame are $\delta_g$ and $\delta_c$, respectively. The voltage phase of SG coincides with $d_g$ axis because of its voltage source characteristics and the current phase of GFLC is determined by the $d_c$-$q_c$ frame because of its current source characteristics. $\varphi_I$ is the angle between the GFLC current and the $d_c$ axis. In this paper, it is a fixed parameter.

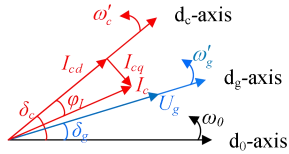

**Fig.2.** Phasor diagram.

### A. Swing Equation-Based LIS Model

Based on the above assumptions, the system is a controlled voltage source. The voltage amplitude is fixed, and the angular dynamic is described by the swing equation of the SG, which can be expressed as:

$$\begin{cases} \dfrac{d\delta_g}{dt} = \omega_g \\ J_g \dfrac{d\omega_g}{dt} = P_{ma,g} - P_{el,g} - D_g\omega_g \end{cases} \tag{1}$$

where $J_g$ and $D_g$ are the inertia and damping coefficient of the SG, respectively. $P_{ma,g}$ is the mechanical power input. $\omega_g = \omega_g^t - \omega_0$. Suppose the output of SG governor is constant during the transient process, $P_{ma,g}$ is a constant. $P_{el,g}$ is the electrical power output of SG, and it can be calculated from the power flow equation:

$$\left.\begin{array}{l} \dot{U}_l = \dot{U}_g - \dot{I}_g\dot{Z}_g \\ \dot{U}_l = \left(\dot{I}_g + \dot{I}_c\right)\dot{Z}_l \end{array}\right\} \Rightarrow \dot{I}_g = \dot{Z}_{eq1}\dot{U}_g + \dot{Z}_{eq2}\dot{I}_c \tag{2}$$

where:

$$Z_{eq1}\angle\theta_1 = \frac{1}{\dot{Z}_g + \dot{Z}_l'}, \quad Z_{eq2}\angle\theta_2 = \frac{\dot{Z}_l'}{\dot{Z}_g + \dot{Z}_l'} \tag{3}$$

where $U_l$ is the voltage of the load node in Fig.1. $I_g$ is the current output of the SG. $Z_g$ is the impedance between SG and the load bus. $Z_l'$ is a virtual impedance. In pre-fault and post-fault period, $Z_l' = Z_l$, and in the fault-on period, $Z_l' = Z_l // R_f$. $Z_{eq1}$ and $Z_{eq2}$ are two virtual impedances to simplify the equation, and the expressions are also given. $\theta_1$ and $\theta_2$ are the phase angle of $Z_{eq1}$ and $Z_{eq2}$, respectively. The upper dot mark indicates that the variable is a vector, and the vector variable without the upper dot mark is the amplitude.

Taking the $d_g$ axis as the reference phase, $P_{el,g}$ can be calculated from the following equation:

$$\begin{aligned} P_{el,g} &= \mathrm{Re}\left(\dot{U}_g\dot{I}_g^*\right) = \mathrm{Re}\left(\dot{U}_g\dot{I}_g\right) \\ &= Z_{eq1}U_g^2\cos\theta_1 - Z_{eq2}U_gI_c\cos\left(\theta_2 + \delta_c - \delta_g + \varphi_I\right) \end{aligned} \tag{4}$$

where the asterisk $^*$ indicates the conjugation of the variable.

### B. PLL Dynamic-Based GFLC Model

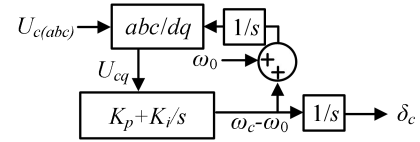

**Fig.3.** PLL diagram.

The GFLC can be regarded as a controlled current source with fixed current amplitude, and the angular dynamics are determined by the PLL dynamics. Fig.3 is a typical diagram of PLL, which adjusts the current angle by the $q_c$-axis voltage of the point of common coupling (PCC). The $q_c$-axis voltage can be calculated from the following equations:

$$\left.\begin{array}{l} \dot{U}_l = \dot{U}_g - \dot{I}_g\dot{Z}_g \\ \dot{U}_l = \left(\dot{I}_g + \dot{I}_c\right)\dot{Z}_l \\ \dot{U}_c = \dot{U}_l + \dot{Z}_c\dot{I}_c \end{array}\right\} \Rightarrow \dot{U}_c = \dot{Z}_{eq2}\dot{U}_g + \dot{Z}_{eq3}\dot{I}_c \tag{5}$$

where:

$$Z_{eq3}\angle\theta_3 = \frac{\dot{Z}_l'\dot{Z}_g}{\dot{Z}_l' + \dot{Z}_g} + \dot{Z}_c \tag{6}$$

$$U_{cq} = Z_{eq2} U_g \sin\left(\theta_2 + \delta_g - \delta_c\right) + Z_{eq3} I_c \sin\left(\theta_3 + \varphi_I\right) \quad (7)$$

where $U_c$ is the voltage of the PCC, and $U_{cq}$ is the q$_c$-axis component of $U_c$. $Z_{eq3}$ is another virtual impedance, and $\theta_3$ is its phase angle.

With $U_{cq}$, the PLL-based phase angle dynamics of GFLC can be expressed as:

$$
\begin{cases}
\dfrac{d\delta_c}{dt} = \omega_c \\[2mm]
\dfrac{d\omega_c}{dt} = K_i U_{cq} + K_p \dfrac{dU_{cq}}{dt} \\[2mm]
\qquad = K_i Z_{eq3} I_c \sin\left(\theta_3 + \varphi_I\right) - K_i Z_{eq2} U_g \sin\left(\delta_c - \theta_2 - \delta_g\right) \\[2mm]
\qquad + K_p Z_{eq2} U_g \cos\left(\theta_2 + \delta_g - \delta_c\right)\left(\omega_g - \omega_c\right)
\end{cases}
\quad (8)
$$

where $K_i$ and $K_p$ are the integral and proportional coefficient of PLL. $\omega_c = \omega_c^t - \omega_0$.

From Fig.3, $\omega_c$ is obtained from a PI controller with $U_{cq}$ as the input. At the moment of fault occurrence or clearing, the virtual load impedance $Z_l'$ is changed ($Z_l$ to $Z_l//R_f$ when the fault occurs and $Z_l//R_f$ to $Z_l$ when the fault is cleared), which leads to the mutation of $Z_{eq2}$ and $Z_{eq3}$, and then leads to the mutation of $U_{cq}$. This causes a mutation of $\omega_c$ through the proportional part of PI controller, which cannot be reflected by (8). The mutation of $\omega_c$ can be expressed by:

$$\Delta\omega_{c,t} = K_p\left(U_{cq,t-} - U_{cq,t+}\right) \quad (9)$$

where, $t$ is the time of fault occurrence or clearing, $U_{cq,t-}$ and $U_{cq,t+}$ are the q$_c$-axis components of $U_c$ before and after time $t$, which are calculated by (3), (6) and (7) with different $Z_l'$. Their characteristics can be found in [4],[21], and not further explained in this paper due to space limitations.

### C. GFLC-LIS Model

The system dynamic equations can be obtained by combining (1), (4) and (8). To guarantee a fixed equilibrium point and simplify the equations, the d$_g$ axis is taken as the reference phase, and the variable $\delta_c$ is replaced by $\delta = \delta_c - \delta_g$, the relative angle between d$_g$ axis and d$_c$ axis, since the reference phase has been changed. The system dynamic equations can be expressed as an ordinary differential equation (ODE) model:

$$
F = \begin{cases}
\dfrac{d\omega_g}{dt} = P_{ma,g}' + P_{el,g}' \cos\left(\theta_2 + \delta + \varphi_I\right) - D_g' \omega_g \\[2mm]
\dfrac{d\delta}{dt} = \omega_c - \omega_g \\[2mm]
\dfrac{d\omega}{dt} = P_{ma,c}' - P_{el,c}' \sin\left(\delta + \varphi - \theta_2\right) \\[2mm]
\qquad - \omega D_c' \cos\left(\delta - \theta_2\right) + D_g' \omega_g
\end{cases}
\quad (10)
$$

where, $\omega = \omega_c - \omega_g$. $P_{ma,g}'$, $P_{el,g}'$, $D_g'$, $P_{ma,c}'$, $P_{el,c}'$, $D_c'$, and $\varphi$ are intermediate parameters and they can be expressed as:

$$
\begin{aligned}
&P_{ma,g}' = \frac{1}{J_g}\left(P_{ma,g} - Z_{eq1} U_g^2 \cos\theta_1\right), \\[2mm]
&P_{el,g}' = \frac{1}{J_g} Z_{eq2} I_c, \\[2mm]
&D_g' = \frac{D_g}{J_g}, \\[2mm]
&P_{ma,c}' = K_i Z_{eq3} I_c \sin\left(\theta_3 + \varphi_I\right) - \frac{P_{ma,g}}{J_g}, \\[2mm]
&P_{el,c}' = \frac{Z_{eq2} U_g}{J_g}\sqrt{\left(K_i J_g\right)^2 + I_c^2 - 2K_i J_g I_c \sin\left(2\theta_2 + \varphi_I\right)}, \\[2mm]
&\varphi = \arctan\left(\frac{I_c \cos\left(2\theta_2 + \varphi_I\right)}{K_i J_g - I_c \sin\left(2\theta_2 + \varphi_I\right)}\right), \\[2mm]
&D_c' = \frac{K_p Z_{eq2} U_g}{J_g}
\end{aligned}
\quad (11)
$$

The dynamics of the variables ($\delta$, $\omega$, $\omega_g$) in the transient process are shown in Fig.4. The parameters in (10) and (11) are given in Table I. The reason for $\omega$ mutation in Fig.4 (b) is the mutation of PLL output. From Fig.4, the system is stable when the fault duration is less than CCT, and the transient analysis in this paper is to provide an accurate estimation of CCT. Since only when the fault clearance time is large than CCT will desynchronization occurs, in practical application, conservative estimation of CCT is acceptable, but radical estimation is unacceptable.

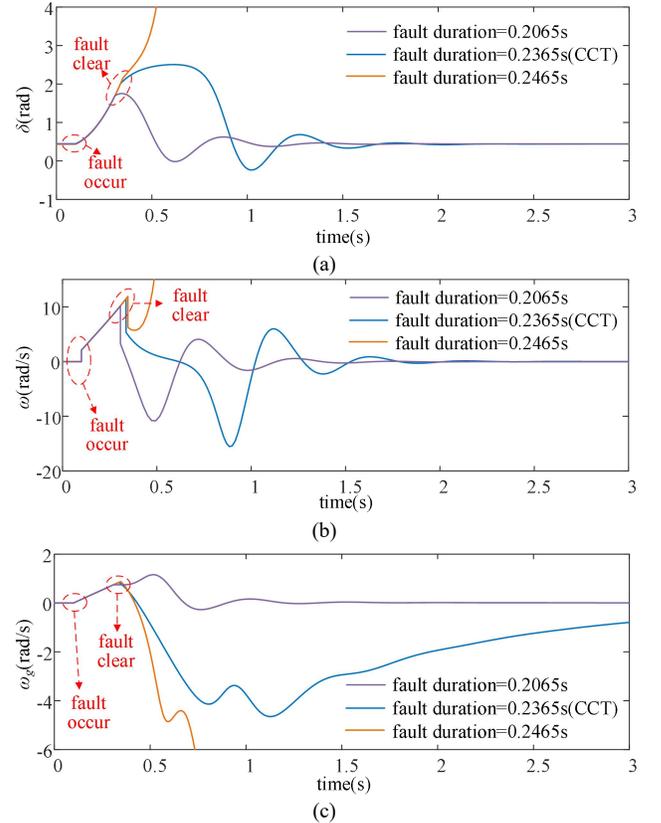

**Fig.4.** Dynamics of the variables ($\delta$, $\omega$, $\omega_g$) in the transient process.

## III. Transient Analysis of GFLC-LIS Based on Zubov Method

### A. Inadaptability of Traditional Transient Analysis Method

From (1), (4) and (8), the dynamic equations of GFLC-LIS are similar to that of double-SG system, but the form of energy function for the double SG system is not applicable to GFLC-LIS. Three reasons are given as follows:

1) By taking the power angle of one SG as the reference phase, its angular velocity can also be eliminated in the dynamic equations, and the variable number of the dual SG system can be reduced to two. The energy function is constructed from the simplified dynamic equations. However, this simplification cannot eliminate the angular velocity of the reference SG in GFLC-LIS, which means there are 3 variables and 3 dynamic equations. The traditional energy function construction method is infeasible.

2) By neglecting $D_g$, the first equation and variable $\omega_g$ in (10) can be eliminated, and the above problem 1) can be solved at the cost of a certain error. However, to construct the energy function from the perspective of virtual kinetic energy, potential energy, and damping energy consumption, it should be ensured that the system damping in the stability region is always positive. It is obvious that when $\delta > \pi/2$, the damping coefficient $D'_c \cos \delta$ in (10) is negative. This leads to a radical CCT estimation, which is unacceptable in transient analysis.

3) By neglecting $D_g$, EAC is a feasible method under negative damping. However in (10), the damping is related to the power angle $\delta$. To calculate the energy consumption of damping, the trajectory of $\delta$ during the on-fault and post-fault process is required, which makes the transient analysis as time-consuming as the simulation method.

The detailed elaboration of the reasons 2) and 3) can be found in the Appendix.

### B. Construction of Energy Function by Zubov Method

In the Zubov method, the energy function $V$ is in the form of an infinite series, which is expressed as:

$$V = V_2 + V_3 + \ldots + V_m + \ldots \tag{12}$$

where, $V_m$ is a polynomial, and degrees of all terms in it are $m$. For practical calculation, $V$ is truncated up to a degree. Let $V^{(M)}$ represent the truncated $V$ at degree $M$.

In a GFLC-LIS, there are 3 variables, and $V_m$ can be expressed as:

$$V_m = \sum_{i=0}^{m} \sum_{j=0}^{m-i} a_{mji} \delta^{m-i-j} \omega^j \omega_g^i \quad m = 2,3,4,\ldots \tag{13}$$

where, $a_{mji}$ is the coefficient of a term in $V_m$, and the degree of $\delta$, $\omega$, $\omega_g$ in this term are $m-i-j$, $j$, and $i$, respectively.

In the Zubov method, the energy function is obtained from the following partial differential equation:

$$\dot{V} = \sum_{n=1}^{N} \frac{\partial V}{\partial x_n} F_n = -\phi(1-V) \tag{14}$$

where, $x$ is the system variable, and $N$ is the dimension. In GFLC-LIS, $(x_1, x_2, x_3) = (\delta, \omega, \omega_g)$, and $N=3$. $\phi$ is a positive defined function of the variables $x$. $F_n$ represents the $n^{th}$ equation in $F$.

**Theorem 1:** Let $A$ represent the attraction domain. For each point $(\delta_0, \omega_0, \omega_{g0}) \in A$, $0 \le V(\delta_0, \omega_0, \omega_{g0}) < 1$. From (13) and (14), it can be concluded that for any $(\delta_0, \omega_0, \omega_{g0}) \in A$, $\dot{V} < 0$.

**Proof:** From (14), one can have:

$$\frac{\partial V}{\partial t} \frac{1}{(1-V)} = -\phi \tag{15}$$

Integrate both sides of (15) to time, and take the logarithm of both sides, one can have:

$$V(\delta_0, \omega_0, \omega_{g0}) = 1 - (1-V) \exp(-J(t)) \tag{16}$$

$$J(t) = \int_0^t \phi \, dt \tag{17}$$

For any $(\delta_0, \omega_0, \omega_{g0}) \in A$, if $t \to \infty$, the state point $(\delta, \omega, \omega_g)$ will converge to the equilibrium point $(0,0,0)$, and $V=0$. Let $t \to \infty$, one can have:

$$V(\delta_0, \omega_0, \omega_{g0}) = 1 - \exp(-J(\infty)) \tag{18}$$

Since $\phi$ is a positive defined function, for all $(\delta_0, \omega_0, \omega_{g0}) \neq 0$, $J(\infty) > 0$. Therefore, from (18), $0 \le V(\delta_0, \omega_0, \omega_{g0}) < 1$.

It is also concluded in [27][28] that the boundary of $A$ is a family curve of $V=1$. The function $\phi$ can influence the solution of $V$, but the boundary of $A$ is not affected.

For a GFLC-LIS, the dynamic equation $F$ in (10) contains non-monomial terms like trigonometric terms. To get the series-form energy function $V$ from (14), Taylor expansions to $F_1$ and $F_3$ are required:

$$\frac{d\omega_g}{dt} = -D'_g \omega_g + P'_{el,I} \cos(2\theta_2 + \varphi_I) \sum_{i=0}^{\infty} \frac{(-1)^i}{2i} x^{2i}$$
$$- P'_{el,g} \sin(2\theta_2 + \varphi_I) \sum_{i=0}^{\infty} \frac{(-1)^i}{2i+1} x^{2i+1} + P'_{ma,g} \tag{19}$$

$$\frac{d\omega}{dt} = D'_g \omega_g - P'_{el,c} \cos \varphi \sum_{i=0}^{\infty} \frac{(-1)^i}{2i+1} x^{2i+1}$$
$$- \left( P'_{el,c} \sin \varphi + \omega D'_c \right) \sum_{i=0}^{\infty} \frac{(-1)^i}{2i} x^{2i} + P'_{ma,c} \tag{20}$$

Let $F'$ represent the Taylor expansion form of $F$, and in this paper, $F'$ is truncated at $M_T=30$. By dividing the linearized and non-linearized parts, $F'$ can also be written as:

$$F'_n = \frac{dx_n}{dt} = \sum_{i=1}^{N} b_{ni} x_i + \sum_{m_T=2}^{M_T} F'_{n,m_T} \tag{21}$$

where, $b_{ni}$ is the coefficient of linearized term $x_i$ in $F'_n$. $F'_{n,m_T}$ contains all $m_T$-degree terms of $F'_n$. Substitute (21) into (14), and compare the coefficients of the terms in both sides of equation (14). $V_2$, $V_3$, ..., $V_m$ can be calculated by:

$$\sum_{n=1}^{N} \frac{\partial V_2}{\partial x_n} \left( \sum_{k=1}^{N} b_{nk} x_k \right) = -\phi \tag{22}$$

$$\sum_{n=1}^{N} \frac{\partial V_m}{\partial x_n} \left( \sum_{k=1}^{N} b_{nk} x_k \right) = R_m(x) \quad m = 3,4,5,\ldots,M \tag{23}$$

where, $R_m(x)$ is a polynomial, each term of which is $m$-degree. Coefficients of its terms are calculated from $V_2$, $V_3$, ..., $V_{m-1}$, and $F'$ by equation (13), and (21). More specifically, $R_m(x)$ is composed of two parts: $\phi V_{m-2}$ from the right side of (14) and $\sum_{i=1}^{m-2} \frac{\partial V_{m-i}}{\partial x_n} F'_{n,i+1}$ from the left side of (14). So, the calculation of $V_m$ is a recursion process, and for each $m$, $V_m$ is obtained from the results of previous stages.

### C. Transient Analysis Based on Zubov Method

Due to the truncation, the set $V^{(M)}=1$ is no longer the boundary of $A$. A conservative approximation of the boundary of $A$ is given in [27]:

Let $W_M$ represent the set of all points other than SEP for which $dV^{(M)}/dt = 0$. For the $V^{(M)}$ values of the points in $W_M$, the maximum and minimum values are represented by $c_2^M$ and $c_1^M$, respectively.

***Theorem 2:*** The set $V^{(M)}=c_1^M$ is wholly contained in $A$.

***Proof:***

1) To prove *Theorem 2*, it is necessary to prove that $dV^{(M)}/dt$ is negative definite inside $V^{(M)}=c_1^M$.

2) Since $(0,0,0)$ is the equilibrium point, it should be guaranteed that the equation $|\lambda I - b|=0$ must have nonzero negative real parts, where $I$ is identity matrix and $b$ is the parameter matrix in (21).

3) In a sufficiently small neighborhood of equilibrium point $(0, 0, 0)$, we have:

$$\frac{dV^{(M)}}{dt} \approx \sum_{m=2}^{M} \sum_{n=1}^{N} \frac{\partial V_m}{\partial x_n} \left( \sum_{k=1}^{N} b_{nk} x_k \right) \qquad (24)$$

$$-\phi + \sum_{m=1}^{M} R_m < 0 \qquad (25)$$

This means $dV^{(M)}/dt$ is negative in a sufficiently small neighborhood of the equilibrium point $(0, 0, 0)$.

4) So, inside $V^{(M)}=c_1^M$, if $dV^{(M)}/dt$ becomes positive, $dV^{(M)}/dt$ will path through zero at some point. Then, this point is in the set $W_M$ with a smaller $V^{(M)}$ value than $c_1^M$, which conflicts with the definition. Therefore, $dV^{(M)}/dt$ is negative definite inside $V^{(M)}=c_1^M$, and *Theorem 2* is proved. More details can be found in [27].

From *Theorem 2*, the proposed transient analysis method is summarized as follows, and the flow chart is shown in Fig.5:

Step 1: obtain the dynamic equations and adopt Taylor expansion to get the series-form of them.

Step 2: calculate the truncated energy function $V^{(M)}$ from (22) and (23).

Step 3: get the set of $dV^{(M)}/dt = 0$, that is, to get $W_M$.

Step 4: get $c_1^M$ in $W_M$. Take the set $V^{(M)}=c_1^M$ as the estimated attraction domain.

Step 5: implement numerical simulation of the on-fault system. At each integration step, calculate the post-fault initial state $(\delta_0^p, \omega_0^p, \omega_{g0}^p)$ from (9). Substitute $(\delta_0^p, \omega_0^p, \omega_{g0}^p)$ to $V^{(M)}$, and the last integration time before $V^{(M)}(\delta_0^p, \omega_0^p, \omega_{g0}^p) > c_1^M$ is the estimated CCT.

In *Theorem 2*, it is guaranteed that $A'$ is conservative, and in actual application, a large $M$ can make the boundary of $A'$ close to $A$. It should also be clear that the boundary of $A'$ does not approach the boundary of $A$ monotonically with the increase of $M$. An $M$ around 16 can achieve high accuracy [28]. The best value of $M$ for the case study system will be discussed in subsequent sections.

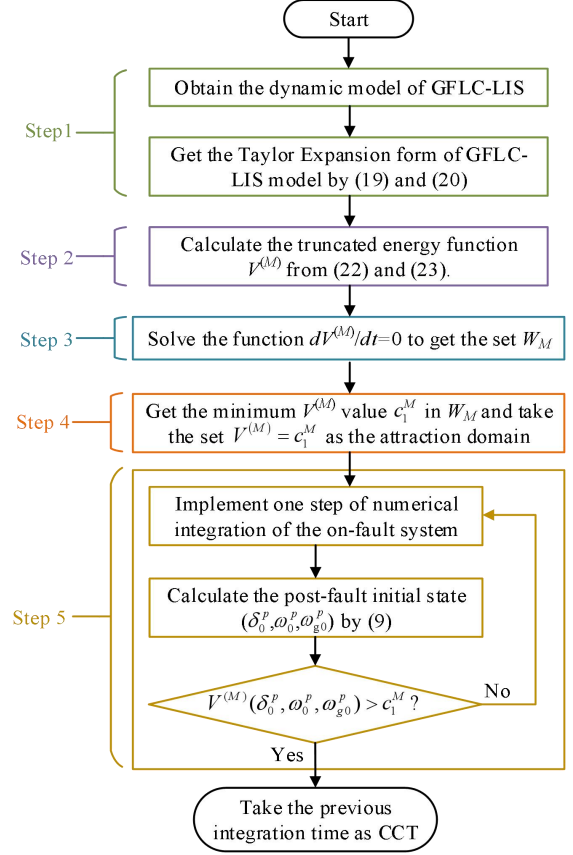

**Fig.5.** Flow chart of the proposed transient analysis method.

From the Zubov-based transient stability analysis process, the pros and cons compared to the traditional Lyapunov method and EAC method can be summarized as:

Cons: 1) Although the Zubov method has been rigorously proven feasible, it does not provide an intuitive explanation from an energy perspective, like a conventional energy function or the EAC method. 2) Recursion is introduced in Step 2, and numerical approaches are introduced in Step 3 and Step 4, which makes the calculation more complicated. However, it still maintains significant computational time advantages compared to the simulation method. 3) Certain parameters, such as $M$, require iterative testing to achieve the highest evaluation accuracy, but decent precision can still be obtained under experience-based parameter selection.

Pros: From the process in Fig.5, the Zubov method imposes less stringent requirements on dynamic equations compared to conventional approaches. More specifically, it remains applicable as long as the dynamic equations can be transformed into polynomial form via Taylor expansion (e.g., (21) ). In contrast, both conventional methods are impractical when path-dependent terms exist (e.g., the negative damping term in (10) ), and the EAC method is restricted to a two-dimensional dynamic function. More details can be found in Section III.A and Appendix.

With the transient stability analysis process in Fig.5, synchronous stability of the GFLC under fixed fault-clearing time can be verified, while control parameters can also be adjusted based on fault-clearing time. Researches show that relatively large PLL parameters ($K_p$, $K_i$) can accelerate the resynchronization process after disturbance, and a relatively large $K_p$ can provide adequate damping [6]. However, too large $K_p$ parameters can lead to a larger phase deviation $\delta_c$ during the fault due to the fast dynamics, which may deteriorate the transient stability. A typical PLL parameter adjustment process based on the Zubov-based stability analysis is shown as follows:

Step 1: Set the original PLL parameter and target CCT. For instance, if the fault clearing time is 0.1s, i.e., 5 cycles, set 0.2s as the target CCT to maintain a certain margin.

Step 2: Get the attraction domain by the Zubov-based method.

Step 3: Implement the on-fault simulation and get the CCT.

Step 4: Compare the CCT with the target CCT. Increase the $K_p$ and $K_i$ and go back to Step 2 until the CCT is close enough to the target CCT.

Each of the above parameter adjustments requires one transient analysis process and one on-fault simulation. However, under the traditional simulation-based method, calculation of CCT requires 5 to 10 times of on-fault and post-fault simulations. Considering the simulation time of on-fault simulation is generally within 0.5s, and the post-fault simulation time is several seconds, the simulation runtime of post-fault simulation is close to 10 times that of on-fault simulation, and each runtime of parameter adjustment under the simulation-based method can be significantly larger than that of the above process.

## IV. CASE STUDY

The transient analysis method in Section III is implemented in GFLC-LIS. The circuit diagram is given in Fig.1, and the parameters are listed in Table I. In particular, the parameters $I_c$ and $\varphi_I$ correspond to converter active and reactive power outputs of 300MW and 50MVar, respectively. Use $\delta_{sep}$ represent the value of $\delta$ at the equilibrium point. In this section, the variable $\delta$ in (10) are replaced by $\delta + \delta_{sep}$, to guarantee that the equilibrium point is always the origin.

### TABLE I
PARAMETERS OF THE TEST SYSTEM

| Parameter | Description | Value |
|---|---|---|
| $S_b$, $U_b$, $\omega_0$ | Base value of system power, voltage, and angular velocity | 100MVA, 230kV,120π rad/s |
| $R_c$, $L_c$ | Converter side system resistance and inductance | 8.928Ω, 0.113H |
| $R_g$, $L_g$ | Grid side system resistance and inductance | 10.631Ω, 0.122H |
| $R_l$, $L_l$ | Load resistance and inductance | 75.571Ω, 7.016H |
| $R_f$ | Fault resistance | 1Ω |
| $S_g$, $S_c$ | Capcity of SG and GFL converter | 400MVA, 400MVA |
| $P_{mag}$, $U_g$ | Mechanical power input and terminal voltage of SG | 366.56MW, 1.1p.u. |
| $J_g$, $D_g$ | Inertia and damping coefficient of SG | 0.4s, 0.4 |
| $K_i$, $K_p$ | Integral and proportional coefficient of PLL | 200, 10 |
| $I_c$, $\varphi_I$ | Magnitude and phase angle of the converter | 0.760p.u., -0.165rad |

### A. Attraction Domain Estimation and ODE Model-Based Transient Stability Analysis

To verify the correctness of the proposed method, the transient analysis process in Section III.C is implemented. The numerical integration of the on-fault system in Step 5 of the transient analysis is firstly based on the ODE model in (10). Taylor expansion form of dynamic equations $F'$ are truncated at $M_f$=30, and the energy function $V$ is truncated at $M$=16.

Firstly, the accuracy of the Taylor expansion is verified by comparing the dynamics of $(\delta, \omega, \omega_g)$ in the transient process under two ODEs: the original ODE in (10) and the Taylor expansion form ODE in (21). The comparison is shown in Fig.6. Results show that there is only a slight deviation between the two ODEs, and the accuracy of the Taylor expansion can meet the application requirement.

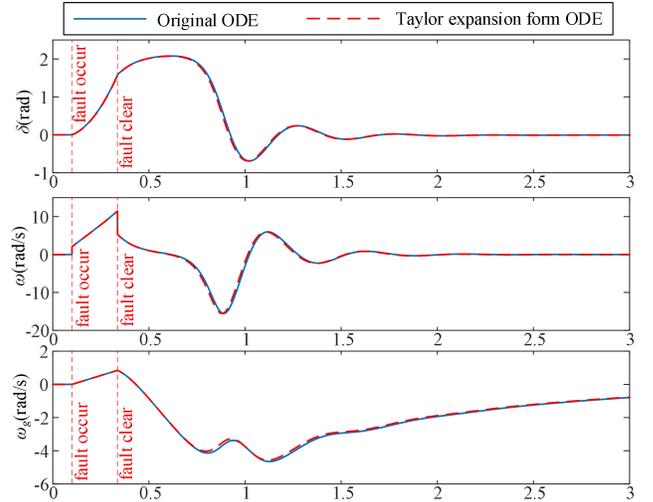

**Fig.6.** Comparison of the original ODE and the Taylor expansion form ODE

Due to the large number of terms in $V^{(16)}$, for example, there are 153 terms in $V_{16}$; only coefficients of $V_2$, and $V_3$ are given in Table II. The coefficients are named according to (13).

### TABLE II
COEFFICIENTS OF ENERGY FUNCTION

| $a_{200}$ | $a_{201}$ | $a_{202}$ | $a_{210}$ | $a_{211}$ | $a_{220}$ |
|---|---|---|---|---|---|
| 0.341054 | 0.002272 | 0.002128 | 0.021873 | 0.003045 | 0.016522 |
| $a_{300}$ | $a_{301}$ | $a_{302}$ | $a_{303}$ | $a_{310}$ | $a_{311}$ |
| -0.03167 | -0.00260 | 0.0002535 | 0.00001114 | -0.02012 | -0.00101 |
| $a_{312}$ | $a_{320}$ | $a_{321}$ | $a_{330}$ | | |
| -0.00006 | 0.000050 | -0.000007 | -0.0000024 | | |

After getting $V^{(16)}$ and $\dot{V}^{(16)}$, the sets $\dot{V}^{(16)}=0$ and $V^{(16)}=c_1^{16}$ are given in Fig.7. From *Theorem 2*, the set $V^{(16)}=c_1^{16}$ can be regarded as a conservative estimation of $A$. To test the error of the proposed method, the transient trajectory under the real CCT and estimated CCT are also given in Fig.7. The fault trajectories are got from the ODE model (10) . The fault occurrence and clear frequency mutations of $\omega$ can be observed on the trajectory. If the post-fault initial state $(\delta_0^p, \omega_0^p, \omega_{g0}^p)$ after fault clear frequency mutation is within $A'$, the system is estimated as stable. It can be seen that the estimated exit point is close to the real exit point. The estimated CCT is 0.2295s, which has a 2.95% error compared with the real CCT of 0.2365s. This is acceptable for practical engineering. The post-fault trajectory under estimated CCT is all within $A'$, which meets the requirements of the attraction domain.

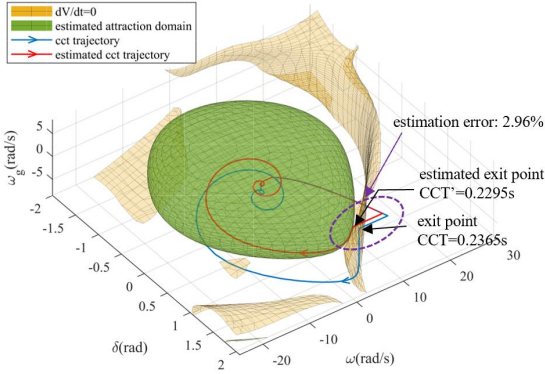

**Fig.7.** Attraction domain estimation and key fault trajectories.

The influence of $M$ on estimation accuracy is analyzed in Fig.8 (a). The estimated 3-dimensional attraction domains are transected at $\omega_g=0$, to visually compare the influence of value $M$ on estimation precision. From Fig.8 (a), $A'$ expands as $M$ increases, and when $M$ exceeds 16, the expansion slows down, and the calculation time increases rapidly. So, $M=16$ is used in this paper. It is worth noting that a larger $M$ does not necessarily correspond to a less conservative estimation. For example, Fig.8 (b), gives part of the estimated boundary with different $M$, and when $M=18$, part of the boundary is less conservative than that when $M=20$. The attraction domains with different $M$ are also given under different LIS or GFLC parameters in the Appendix, and similar conclusions can be drawn.

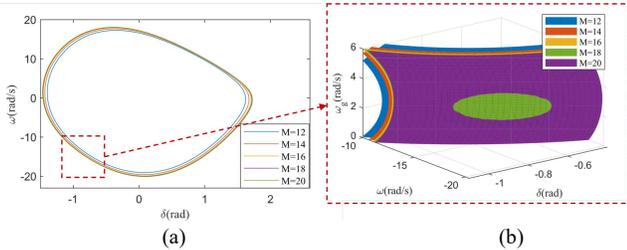

**Fig.8.** Comparison of attraction domain with different $M$.

In Fig.7, the equation $\phi$ is :

$$\phi_1 = 0.03*\delta^2 + 0.03*\omega^2 + 0.03*\omega_g^2 \tag{26}$$

The equation $\phi$ is obviously positive defined, and the coefficients are set according to the range of the variables, generally avoiding the terms larger than 10 [28].

Other form of the equation $\phi$ are also tested, which are:

$$\phi_2 = 0.01*(\delta + \omega + \omega_g)^2 \tag{27}$$

$$\phi_3 = 0.03*\delta^2 + 0.01*(\omega + \omega_g)^2 \tag{28}$$

The estimated CCT under $\phi_2$ and $\phi_3$ are 0.2289s and 0.2292s, which are all close to the result under $\phi_1$. The estimated attraction domains under different equations $\phi$ are shown in Fig.9. The estimation domains are very close, indicating that the choice of equation $\phi$ does not have a prominent influence on the transient stability analysis.

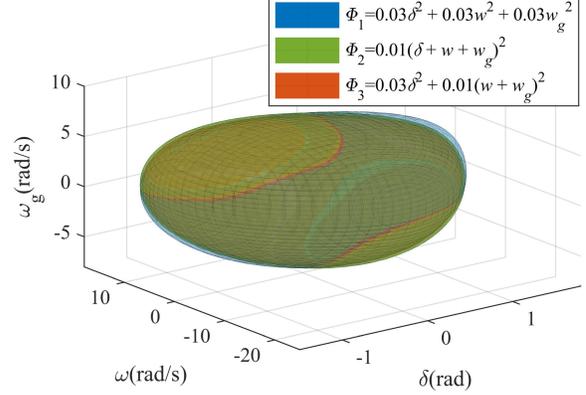

**Fig.9.** Comparison of the attraction domain with different LIS parameters.

It should be noticed that in Steps 3 and 4 of Fig.5, numerical approaches are required due to the nonlinearity of the function $dV^{(M)}/dt$ . Whether the errors brought by the numerical approaches will influence the transient analysis and CCT estimation needs further verification. Under the parameters in Table 1, $c_1^M =0.572$. Supposing the numerical approaches lead to a $\pm 1\%$ error of the $c_1^M$ (which is a relatively large error for numerical approaches), the actual $c_1^M$ can be in the range of (0.566, 0.578). Taking the boundary $V^{(16)} = 0.566$ and $V^{(16)} = 0.578$ as the stability boundary, the corresponding estimated CCTs are 0.2289s and 0.2302s, which are very close to the original estimated CCT of 0.2295s. Considering the actual CCT 0.2365s, the conservativeness of the Zubov method is not changed. The boundaries under different $c_1^M$ conditions and the fault trajectories under corresponding CCTs and real CCTs are shown below.

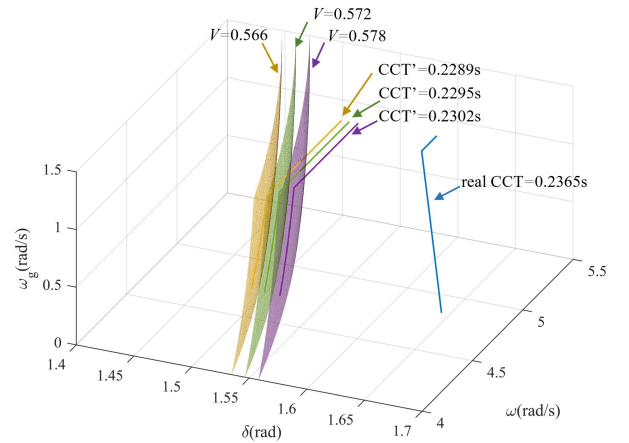

**Fig.10.** Attraction domain estimation and key fault trajectories considering the error of numerical approaches.

A comparison between the traditional energy function and the Zubov-based energy function is shown in Fig.11. The traditional energy function is given in the Appendix. From Fig.11, two problems of the traditional energy function can be observed: Firstly, the influence of $\omega_g$ on the transient stability is neglected, which obviously influences the estimation accuracy. Secondly, radical error is introduced. The unstable trajectory corresponds to the fault-clearing time of 0.237s, which is larger than the real CCT. However, it can be observed that the post-fault initial state $(\delta_0^p, \omega_0^p, \omega_{g0}^p)$ after fault clear locates inside the estimated attraction domain under the traditional energy function, which means it will be judged as stable, and a radical error is introduced. This is unacceptable for CCT estimation. In contrast, there is only conservative error under the Zubov-based energy function.

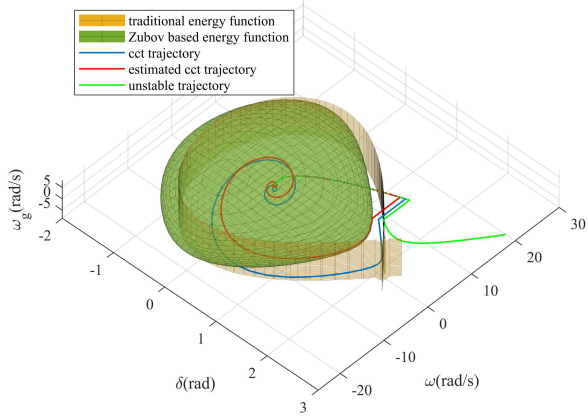

**Fig.11.** Comparison of the attraction domain under two energy functions

Two more cases with different $I_c$ and $\varphi_I$ are tested to evaluate the error of the proposed method. The results of CCT estimation are shown in Table III, and the estimation errors are all below 5%. In Case 2, the converter and SG active power outputs are 350 MW and 314.64 MW, which leads to a higher converter source proportion compared to Case 1. The CCT is decreased, and the transient stability is also weakened. For Case 3, the converter reactive power output is reduced to 20 MVar. The decrease in reactive power support also causes deterioration of transient stability.

TABLE III
ESTIMATED CCT UNDER DIFFERENT POWER OUTPUT

|  | Case 1 | Case 2 | Case 3 |
|---|---|---|---|
| Converter current | $I_c$=0.760p.u. $\varphi_I$=-0.165rad | $I_c$=0.884p.u. $\varphi_I$=-0.142rad | $I_c$=0.752p.u. $\varphi_I$=-0.067rad |
| Estimated CCT' | 0.2295s | 0.1751s | 0.2132s |
| Real CCT | 0.2365s | 0.1842s | 0.2229s |
| Error | 2.96% | 4.94% | 4.35% |

The influence of inertia and damping of SG (low-inertia system) on attraction domain is also analyzed, and the estimated attraction domains with different inertia and damping are shown in Fig.12 (a) and (b). From Fig.12 (a), with the increase of inertia $J_g$, $\mathbf{A'}$ shrinks in $\omega_g$ dimension and expands in $\omega$ and $\delta$ dimension, which means the system can ride through larger disturbance of $\omega$ and $\delta$, but smaller disturbance of $\omega_g$. A larger inertia decelerates the dynamic of the LIS, which makes it easier for the VSC to synchronise but harder to maintain stability when a certain $\omega_g$ disturbance occurs. From Fig.12 (b), with the increase of damping $D_g$, $\mathbf{A'}$ expands in all dimensions, which means the system can ride through larger disturbance of $\omega$, $\omega_g$ and $\delta$.

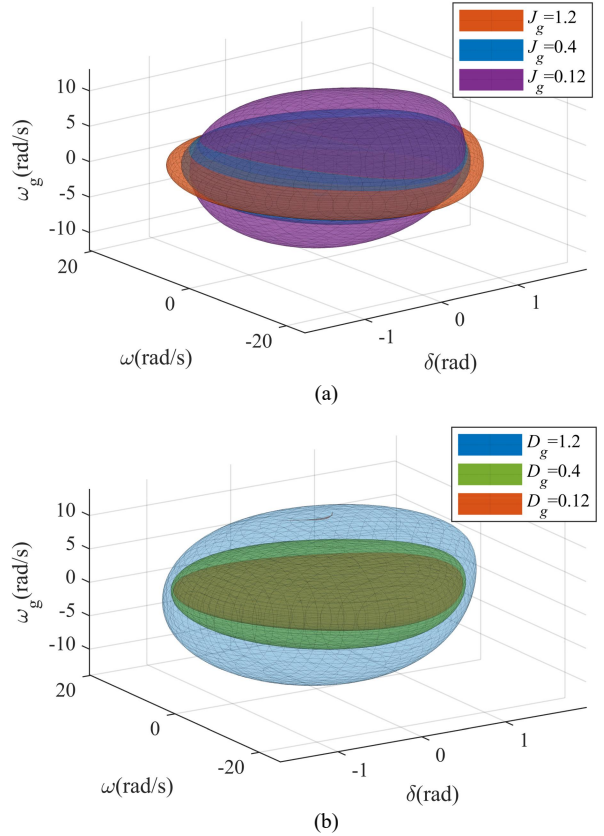

(a)

(b)

**Fig.12.** Comparison of attraction domain with different LIS parameters.

A large attraction domain doesn't always correspond to a large CCT. The fault-on trajectory to the stability boundary and the movement speed of the state point on the trajectory are also key influences of CCT. In Table IV, CCTs under different inertia and fault resistance are compared. When fault resistance $R_f$ is low, CCT decreases with the increase of inertia, and in large fault resistance $R_f$ cases, CCT increases with the increase of inertia. CCTs under different damping and fault resistance are given in Table V, and the results are similar to those in Table IV. CCT is negatively correlated to damping in low $R_f$ cases and positively correlated to damping in large $R_f$ cases. This is inconsistent with the general belief that the larger the inertia or damping, the better the transient stability.

TABLE IV
CCT UNDER DIFFERENT INERTIA AND FAULT RESISTANCE

|  | Case 4 | Case 5 | Case 6 | CCT decreases with the increase of inertia |
|---|---|---|---|---|
| Inertia | $J_g$=1.2p.u. | $J_g$=0.4p.u. | $J_g$=0.12p.u. |  |
| Fault resistance | $R_f$=1$\Omega$ | $R_f$=1$\Omega$ | $R_f$=1$\Omega$ |  |
| Estimated CCT | 0.2266 | 0.2295s | 0.2402 |  |
| Real CCT | 0.2315 | 0.2365s | 0.2511 |  |

| | Case 7 | Case 8 | Case 9 | CCT increases with the increase of inertia |
|---|---|---|---|---|
| Inertia | $J_g$=1.2p.u. | $J_g$=0.4p.u. | $J_g$=0.12p.u. | |
| Fault resistance | $R_f$=10Ω | $R_f$=10Ω | $R_f$=10Ω | |
| Estimated CCT | 0.4442 | 0.3543 | 0.2405 | |
| Real CCT | 0.4648 | 0.3868 | 0.2815 | |

TABLE V
CCT UNDER DIFFERENT DAMPING AND FAULT RESISTANCE

| | Case 10 | Case 11 | Case 12 | CCT decreases with the increase of damping |
|---|---|---|---|---|
| Damping | $D_g$=1.2p.u. | $D_g$=0.4p.u. | $D_g$=0.12p.u. | |
| Fault resistance | $R_f$=1Ω | $R_f$=1Ω | $R_f$=1Ω | |
| Estimated CCT | 0.2274s | 0.2295s | 0.2236s | |
| Real CCT | 0.2347s | 0.2365s | 0.2373s | |

| | Case 13 | Case 14 | Case 16 | CCT increases with the increase of inertia |
|---|---|---|---|---|
| Damping | $D_g$=1.2p.u. | $D_g$=0.4p.u. | $D_g$=0.12p.u. | |
| Fault resistance | $R_f$=10Ω | $R_f$=10Ω | $R_f$=10Ω | |
| Estimated CCT | 0.3924s | 0.3543s | 0.2894s | |
| Real CCT | 0.4125s | 0.3868s | 0.3765s | |

The LIS angular dynamic is the reason for the above relationship between $R_f$ and CCT. When $R_f$=10Ω, during the fault, the phase of PLL accelerates and the phase of SG decelerates, and the opposite dynamics expand phase deviation between the two devices (i.e. $\delta$). In this case, the larger the inertia, the slower the deceleration, which can reduce phase deviation between two devices (i.e., $\delta$), thereby enhancing transient stability.

When $R_f$=1Ω, both PLL phase and SG phase accelerate during the fault, and the dynamic of SG reduces the phase deviation between the two devices (i.e. $\delta$). In this case, the lesser the inertia, the faster the acceleration. Considering the PLL dynamics are always faster than the SG dynamics, a relatively small inertia can help the SG phase to catch up with the PLL phase during the acceleration, which reduces the phase deviation between the two devices (i.e., $\delta$), thereby enhancing transient stability. Dynamics of $\omega_g$ under different $R_f$ are shown in Fig.13. More specific analysis of the acceleration and deceleration of the angular dynamics are given in the Appendix.

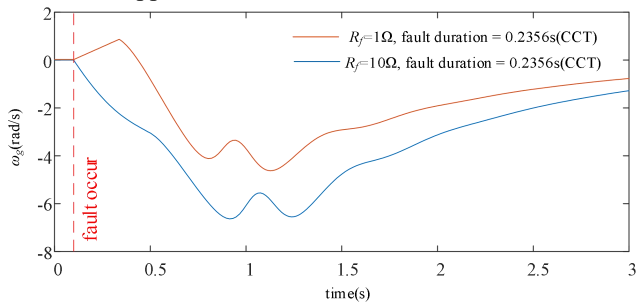

**Fig.13.** Dynamics of $\omega_g$ under different $R_f$.

The influence of PLL dynamic on transient stability is analyzed. The attraction domains with different PLL parameters are shown in Fig.14. Larger $K_i$ and $K_p$ can increase the dynamic speed of PLL, and influence the transient stability of the system. From Fig.14, with the increase of $K_i$ and $K_p$, the attraction domain expands in $\omega$ and $\omega_g$ dimension, but in $\delta$ dimension, the attraction domain first expands and then shrinks. Faster dynamic allows the PLL to realize quicker synchronization when disturbance occurs, therefore enlarge the attraction domain, especially in $\omega$ dimension.

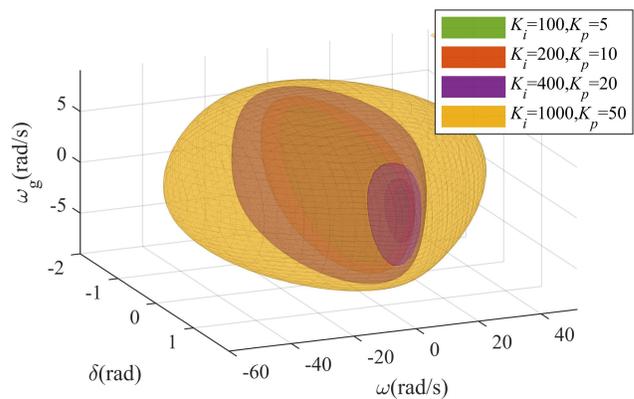

**Fig.14.** Comparison of attraction domain with different PLL parameters.

CCTs under different PLL parameters are compared in Table VI. Though the attraction domain is enlarged with the increase of $K_i$ and $K_p$, CCT decreases. This is because the dynamic of VSC is significantly quicker than that of LIS, which makes the phase angle deviation $\delta$ and angular speed deviation $\omega$ rapidly increase during on-fault period, and when the PLL dynamic is similar to LIS, the increase of $\omega$ and $\delta$ during on-fault period slows down. This can also be reflected by the equations of the intermediate parameters in (11): by setting $K_i$ and $K_p$ properly, parameters like $P'_{el,c}$ can be significantly reduced, which greatly reduce the dynamic of the system and improve the transient stability.

TABLE VI
CCT UNDER DIFFERENT DAMPING AND FAULT RESISTANCE

| | Case 17 | Case 18 | CCT decreases with the increase of PLL dynamic speed |
|---|---|---|---|
| PLL parameters | $K_p$=5 $K_i$=100 | $K_p$=10 $K_i$=200 | |
| Estimated CCT | 0.3425s | 0.2295s | |
| Real CCT | 0.3554s | 0.2365s | |
| | Case 19 | Case 20 | |
| $K_p/K_i$ | $K_p$=20 $K_i$=400 | $K_p$=50 $K_i$=1000 | |
| Estimated CCT | 0.1540s | 0.0841s | |
| Real CCT | 0.1582s | 0.0910s | |

### B. Simulation-Based Transient Stability Analysis

To further test the accuracy of the proposed method, the estimated attraction domain is tested by the simulation results. The simulations of the GFLC-LIS under different conditions are conducted with PSCAD/EMTDC. The diagram of the system is in Fig.1, and the parameters are in Table I. The attraction domain estimation is already implemented in Section IV.A, and the transient analysis process in Section III.C is based on the simulation model rather than the ideal ODE (ordinary differential equation) model, which is carried out in this section.

The simulation results are shown in Fig.15(a). The CCT of the simulation model is 0.224s, which means the ideal ODE model has an error of around 0.01s on CCT. This is because reactance is regarded as a constant, and the influence of frequency fluctuation is neglected. The fault occurrence and clear frequency mutations have a relatively large influence on system dynamics, which is the main source of error. However, according to the transient analysis steps in Section IV.A, the ODE model is not directly used for CCT estimation but only to get the attraction domain of the post-fault system, which is not influenced by the frequency mutation.

The estimated attraction domain, the estimated exit point before and after the fault clearing, and the two fault trajectories in Fig.15 (a) are given in Fig.15 (b), to analyze the accuracy of the attraction domain. Implement the Step 5 of the analysis method in Section III.C. When the post-fault initial state $(\delta_0^p, \omega_0^p, \omega_{g0}^p)$ is first outside the attraction domain (corresponding $V^{(16)}$ value is larger than $c_1^{(16)}$), the former state is regarded as the exit point. The estimated CCT is 0.219s, which has a conservative 2.23% error compared to the real CCT, 0.224s.

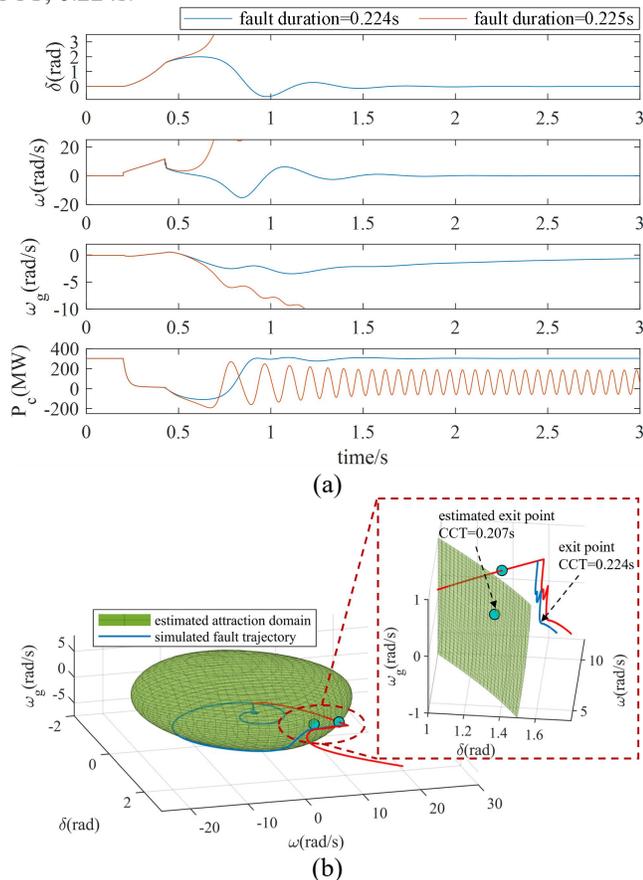

**Fig.15.** Fault trajectories and transient analysis results of GFLC-LIS simulation model.

TABLE VII
ESTIMATED CCT UNDER DIFFERENT POWER OUTPUT

|  | Case 1 | Case 2 | Case 3 |
|---|---|---|---|
| Converter current | $I_c$ =0.760p.u. $\varphi_l$=−0.165rad | $I_c$ =0.884p.u. $\varphi_l$=−0.142rad | $I_c$ =0.752p.u. $\varphi_l$=−0.067rad |
| Estimated CCT' | 0.219s | 0.169s | 0.204s |
| Real CCT | 0.224s | 0.175s | 0.210s |
| Error | 2.23% | 3.43% | 2.86% |

Case 2 and Case 3 in Table III are also tested by the simulation model. The CCT estimation results are in Table VII. With larger active power proportion and less reactive power support, respectively (determined by parameter $I_c$ and $\varphi_l$), the CCT in both cases are reduced, but the CCT estimations of all 3 cases are all conservative and the errors are all less than 5%, which are similar to the results base on ideal ODE in Table III.

The run time of the analysis process is as follows. The calculation of the attraction domain requires 63.3s, and the on-fault simulation requires only 10.6s (simulation time is 0.5s). The calculation of the CCT only requires 73.9s. However, the on-fault and post-fault simulation requires 66.1s (simulation time is 5s), and 8 simulations are required to get the CCT with sub-millisecond precision (<0.001s) by dichotomy (initial fault clearing time is 0.4s). Therefore, the CCT calculation by the simulation-based method is 528.8s, which is significantly larger than that of the proposed Zubov-based method.

## V. CONCLUSION

This paper investigates the transient stability of the GFLC-LIS. Both the angular dynamics of the low-inertia system and the negative damping of GFLC are considered. The inability of traditional methods due to these two problems is illustrated, and a Zubov-based transient analysis method is proposed. The impacts of method parameters and system parameters are also analyzed. The main conclusions are drawn as follows:

1) The feasibility of the proposed Zubov-based method is not influenced by the above two problems. Results show that in most cases, the estimation error is less than 5%, and relatively large errors occur only in large fault resistance, low inertia/damping cases. In contrast, radical error is observed under the traditional Lyapunov method, and the EAC method is not feasible for a 3-dimensional system when the LIS dynamic is considered.

2) Compared to the simulation-based method, the Zubov-based method has a significant advantage in calculation time. This makes it suitable for stability verification under multiple scenarios or giving guidance to the parameter setting of the GFLC.

3) Appropriately increasing the order of the energy function can reduce the conservation of the estimation, and for the cases of this paper, $M$=16 is an appropriate choice considering both accuracy and calculation speed.

## ACKNOWLEDGMENT

This work is supported by the National Natural Science Foundation of China (U22B6007) and Science and Technology Project of China Southern Power Grid Co. Ltd. (ZBKJXM20232419).

## APPENDIX

### A. Inadaptability of Traditional Energy Function and the EAC Method

Firstly, to apply EAC method or establish the traditional energy function from the conservation of energy perspective, the parameter $D_g$ is approximated as zero. Then first equation in (10) can be eliminated, and (10) can be converted to (29).

$$F = \begin{cases} \dfrac{d\delta}{dt} = \omega \\ \dfrac{d\omega}{dt} = P'_{ma,c} - P'_{el,c}\sin\left(\delta + \delta_{sep} + \varphi - \theta_2\right) \\ \qquad - \omega D'_c \cos\left(\delta + \delta_{sep} - \theta_2\right) \end{cases} \quad (29)$$

The dynamic equations in (29) are in the form of the swing equation but incorporate variable damping. In traditional methods, like the traditional Lyapunov method or EAC method, the damping term is generally neglected since it is always positive. However, in (29), the damping term (the last term in the second equation) changes with $\delta$ and can be negative. If the damping term is ignored, the positive definiteness of the energy function is not guaranteed, which leads to radical errors. If we add the damping work into the traditional methods, the post-fault trajectory is required, which causes the traditional methods to lose the advantage of faster speed compared to the simulation method.

To give a comparison of the traditional energy function and the Zubov-based energy function, ignoring the damping term (the last term in the second equation), the traditional energy function is obtained, including virtual kinetic energy and potential energy:

$$V_{tr} = \frac{1}{2}\omega^2 - P'_{ma,c}\delta \\ - P'_{el,c}\left(\cos(\delta + \delta_{sep} + \varphi - \theta_2) - \cos(\delta_{sep} + \varphi - \theta_2)\right) \quad (30)$$

The comparison of the attraction domain under two energy functions is shown in Fig.11.

### B. Analysis of the Parameter M under Different LIS or GFLC

The stability domains under different $M$ are also tested under the system inertia $J_g$=1.2 and the GFLC parameter $K_i$=400, $K_p$=20. The results are shown below. The right figures are the attraction domains, and the figures on the left are the transections of the attraction domains at $\omega_g$=6, $\omega_g$=0, $\omega_g$=-6. Similar results can be obtained compared to the case in Fig. 8. More specifically, the stability domains do not significantly expand when $M$>16, and considering the increasing calculation time, $M$=16 is a decent choice.

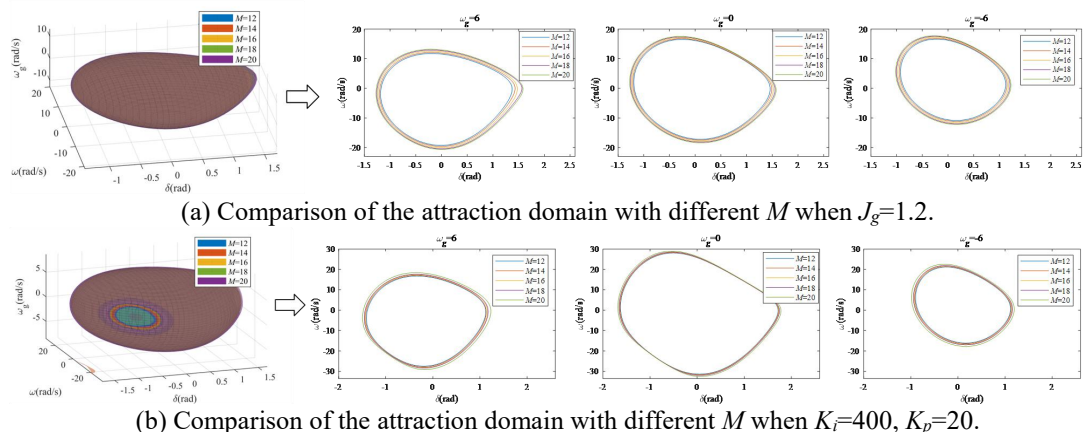

(a) Comparison of the attraction domain with different $M$ when $J_g$=1.2.

(b) Comparison of the attraction domain with different $M$ when $K_i$=400, $K_p$=20.

**Fig.16.** Comparison of the attraction domain with different $M$ under different LIS and GFLC parameters.

### C. Phase Acceleration/Deceleration Analysis

Whether the SG or PLL phase accelerates or decelerates is determined by the ODE. For the SG phase, it is determined by the first function in (10), which is composed of a virtual mechanical power term $P'_{ma,g}$, a virtual electrical power term $P'_{el,g}\cos(\theta_2 + \delta + \varphi_I)$, and a virtual damping term $-D'_g\omega_g$. When the sum of the power terms is negative, $\frac{d\omega_g}{dt}$ is negative, and $\omega_g$ reduces. When the sum of the power terms is positive, $\frac{d\omega_g}{dt}$ is positive, and $\omega_g$ increases. For the PLL phase, it is determined by the second function in (8).

When a fault occurs, the parameters $Z_{eq1}$, $Z_{eq2}$, and Zeq3 change, which in turn affects the parameters in (10) and (8). Taking the SG phase as an example, the parameters $P'_{ma,g}$, $P'_{el,g}$, and $D'_g$ are shown as in the following table. When $R_f$=1Ω, at the instant of fault occurrence, $\omega_g$=0, and the value $P'_{ma,g}$ and $P'_{el,g}$ makes the right side of the first function in (10) positive, and SG phase accelerates. When $R_f$=10Ω, at the instant of fault occurrence, $\omega_g$=0, and the value $P'_{ma,g}$ and $P'_{el,g}$ makes the right side of the first function in (10) negative, and the SG phase decelerates.

TABLE VIII
ODE PARAMETER VALUE UNDER DIFFERENT CONDITIONS

|  | No fault | $R_f$=1Ω | $R_f$=10Ω |
|---|---|---|---|
| $P'_{ma,g}$ | -19.9608 | 3.6105 | -13.2370 |
| $P'_{el,g}$ | 25.6091 | 0.6906 | 5.9128 |
| $D'_g$ | 1 | 1 | 1 |